\documentclass[twocolumn,showpacs,amsmath,amssymb,prd,nofootinbib]
{revtex4}
\usepackage{graphicx}
\def\sss{\scriptscriptstyle}
\def\^#1{^{\sss #1}}
\def\_#1{_{\sss #1}}
\def\beq{\begin{equation}}
\def\eeqno#1{\label{#1}\end{equation}}

\def\rarrow{\rightarrow }

\def\dleft{\rlap{{\it D}}\raise 8pt\hbox{$\scriptscriptstyle\Leftarrow$}}
\def\dright{\rlap{{\it
D}}\raise 8pt\hbox{$\scriptscriptstyle\Rightarrow$}}

\def\cmss{{\rm cm~s^{-2}}}
\def\kpc{{\rm ~kpc}}
\def\Mpc{{\rm ~Mpc}}

\def\az{a\_{0}}

\def\mlsun{(M/L)\_\odot}
\def\l0{\ell\_{0}}

\def\l{\lambda}

\def\m{\mu}
\def\n{\nu}

\def\xlimin{{x\rarrow\infty \atop{\raise 1pt\hbox to 30pt{\rightarrowfill}}}}
\def\limlim#1#2{{#1\rarrow #2 \atop{\raise 1pt\hbox to 30pt{\rightarrowfill}}}}

\def\gN{g\_N}

\begin{document}

\title{Testing MOND Over a Wide Acceleration Range in X-Ray Ellipticals}
\author{Mordehai Milgrom} \affiliation{DPPA, Weizmann Institute of
Science, Rehovot 76100, Israel}

\begin{abstract}
The gravitational fields of two isolated ellipticals, NGC 720 and NGC 1521, have been recently measured to very large galactic radii ($\sim 100$ and $\sim 200 \kpc$), assuming hydrostatic balance of the hot gas enshrouding them. They afford, for the  first time to my knowledge, testing MOND in ellipticals with force and quality that, arguably, approach those of rotation-curve tests in disk galaxies. In the context of MOND, it is noteworthy that the measured accelerations span a wide range, from more than $10\az$ to about $0.1\az$, unprecedented in individual ellipticals. I find that MOND predicts correctly the measured dynamical mass runs (apart from a possible minor tension in the inner few $\kpc$ of NGC 720, which might be due to departure from hydrostatic equilibrium): The predicted mass discrepancy increases outward from none, near the center, to $\sim 10$ at the outermost radii. The implications for the MOND-vs-dark-matter controversy go far beyond the simple fact of two more galaxies conforming to MOND.
\end{abstract}

\pacs{04.50.-h  98.52.Eh  98.80.-k}
\maketitle

\section{introduction}
MOND, which posits a departure from standard dynamics as an alternative to dark matter, has been amply tested in disk galaxies of all types (for a recent review of MOND, see e.g., Ref. \cite{fm12}).
These forceful tests have been made possible by the presence of neutral gas, in ordered, quasi-circular, rather well-understood  motion, observed to large radii. Rotation-curve analysis remains the flagship of MOND testing.
\par
Testing MOND in individual elliptical galaxies is far less advanced.
In most ellipticals our only means of probing the gravitational field is the study of their stellar dynamics. But, this probe suffers serious uncertainties due to our ignorance of the stellar motions (the orbits of stars are not circular and can be anisotropic, with unknown dependence on radius).
For example, Ref. \cite{salinas12} find from the stellar kinematics of NGC 7507, some tension with the predictions of MOND, but they studied only a rather limited set of orbit distributions. In contrast, Sanders (2012, shown in Fig. 37 of Ref. \cite{fm12}) found it easy to get a very good match of the same data with the MOND predictions, for other orbit distributions.
\par
Furthermore, this method allows us to probe the field only to relatively small radii, i.e., within the stellar component.
But ellipticals have, by and large, high mean surface densities. Expressed as accelerations, these correspond to mean accelerations of the order of $\az$ or higher, within the bulk of the stellar component.
MOND then predicts that only very modest mass discrepancies should be found within these regions. (The term ``mass discrepancy'' refers to the ratio of the mass deduced by dynamical means, assuming Newtonian dynamics--the dynamical mass--and the mass detected directly--the baryonic mass.) This prediction is indeed born out by the observations. But it means that we cannot expect to test MOND, using stellar dynamics, in the more interesting regime of low accelerations and large predicted mass discrepancies.
\par
Analysis using strong lensing of quasars avoids the first shortcoming but not the second. The ``coincidence'' $2\pi\az\approx c^2/d\_H$ \cite{milgrom83} ($d\_H$ is the Hubble distance) implies that
strong lensing cannot occur in the low-acceleration regime
for a low-redshift  elliptical and a source at cosmological distances.
So, only very mild mass discrepancies are predicted by MOND to appear in such studies. Also, these studies produce only one, bulk value for the mass discrepancy, not a run with radius. This leaves the interpretation of such analyses at the mercy of knowing accurately the stellar mass-to-light ratios of the elliptical in question. It is not surprising, then, that one finds different groups reaching different conclusions using even the same data. For example, Ref. \cite{ferreras08} claim meaningful (albeit small) tension between the strong-lensing, MOND masses, and the stellar masses they deduce in the central, high-acceleration regions of several ellipticals, while Ref. \cite{chiu08} claim success for the MOND predictions applying the same technique to a larger sample, including that of Ref. \cite{ferreras08} (see also Refs. \cite{chen06,sl08,chiu11}, who reached similar conclusions as Ref. \cite{chiu08}).
\par
In a rare case, the presence of a rotating gas ring around the elliptical NGC 2974 permitted \cite{weijmans08}
to test MOND up to a radius of about $12\kpc$, using the ring's rotation curve. They find good agreement with the predictions of MOND for the interpolating function $\m_1$ (see below). But, again, only small mass discrepancies appear within the region that is probed.
\par
Using more luminous test objects, such as planetary nebulae, or globular clusters, one can probe individual ellipticals to somewhat larger radii. A notable example is the MOND analysis in Ref. \cite{ms03}, of data by Ref. \cite{romanowsky03} on planetary nebulae in three ellipticals, reaching 3-5 effective radii. But these still do not reach very far, where large discrepancies develop, and the method still greatly suffers from large uncertainties due to ignorance of the orbit distribution. For instance, Ref. \cite{schuberth12} used the globular-cluster system of NGC 4636 to probe for dark-matter and MOND. They find consistency with the MOND predictions, but that no strong statement can be made since the accelerations probed are not low enough.
\par
In some ellipticals we can observe x-ray emitting hot gas to relatively large radii. But these tend to be in clusters or groups, and, in particular, to be the central galaxy of the cluster. The mass discrepancies such analyses reveal might then be related to those of the cluster at large.
\par
In light of all this, the recent advent of two studies of isolated ellipticals, probing their dynamics to large radii, is most welcome. The dynamics of NGC 720 \cite{humphrey11} and NGC 1521 \cite{humphrey12} were studied assuming hydrostatic equilibrium in their x-ray emitting gas envelops. The gravitational field was probed to a little under $100 \kpc$ for the former, and a little under $200 \kpc$ for the latter. From the deprojected x-ray temperature and density runs, \cite{humphrey11,humphrey12} deduced the spherically averaged runs of dynamical mass, $M_d(R)$, which is equivalent, conceptually, to measuring the rotation curve, $v(R)=[GM_d(R)/R]^{1/2}$, or the acceleration field $g(R)=v^2(R)/R$.
\par
The accelerations measured, $g$, span a range from more than $10\az$ near the center to about $0.1\az$ at the outer radii. In this regard, they are as probing as the best of rotation-curves.
And, while this method brings its own systematics (discussed extensively by Refs. \cite{humphrey11,humphrey12}--e.g., unaccounted for departures from hydrostatic equilibrium: bulk motion of the gas--such as infall, rotation, and turbulence; uncertainties introduced by having to deproject the gas and stellar distributions, etc.) they afford, like rotation-curves, analysis free of our ignorance of the test-particles orbits.
So, all in all, they permit us, for the first time, as far as I am aware, to test MOND in individual ellipticals, with almost the same effectiveness as in rotation-curve analysis.\footnote{The deduction of runs of dynamical mass, using gas hydrostatics, has, arguably, not yet reached the accuracy of measuring the analogous HI rotation curves in spirals.}$^,$\footnote{Weak-lensing (e.g., Ref. \cite{tian09}), which does probe large radii, does not probe individual galaxies, only some population averages, and is still beset by large errors.}
\par
Here I compare the derived runs of $M_d$ with the detailed MOND predictions for these runs, based on only the observed baryon distribution (for reasonable stellar $M/L$ values).
The analysis and results are described in section \ref{analysis}, and discussed in section \ref{discussion}.

\section{\label{analysis}Analysis and results}
The dynamical analysis of Refs. \cite{humphrey11,humphrey12} approximates the systems as spherical, and so will mine.
For a spherical system, all existing formulations of MOND as modified gravity predict the relation between the measured acceleration, $g$, and the Newtonian acceleration produced by the baryons, $\gN$:
\beq \gN=\mu(g/\az)g. \eeqno{ii}
Here $\m(x)$ is the MOND interpolating function, appearing, e.g., in the modified Poisson theory of Ref. \cite{bm84}.
Its asymptotic behaviors are $\m(x\ll 1)\approx x$, $\m(x\gg 1)\approx 1$. We can also write
\beq g=\n(\gN/\az)\gN, \eeqno{iii}
$\nu(y)$ is the interpolating function appearing, more directly, in QUMOND \cite{milgrom10}. The two theories are equivalent for spherical systems if the two interpolating functions are related by $\m(x)=1/\n[x\m(x)]$.
\par
The dynamical mass enclosed within radius $R$ is $M_d(R)\equiv G^{-1}g(R)R^2$, while $\gN(R)=GM_b(R)/R^2$, where $M_b(R)$ is the enclosed baryonic mass.
Those MOND formulations (and the relativistic theories that reduce to them in the nonrelativistic limit) thus predict that in isolated spherical systems, $M_d(R)$ and $M_b(R)$ are related by\footnote{In the dark-matter paradigm, the two are not related.}
 \beq M_d(R)=\n\left[\frac{GM_b(R)}{R^2\az}\right]M_b(R). \eeqno{iv}
\par
\begin{figure}
\begin{center}
\includegraphics[width=0.95\columnwidth]{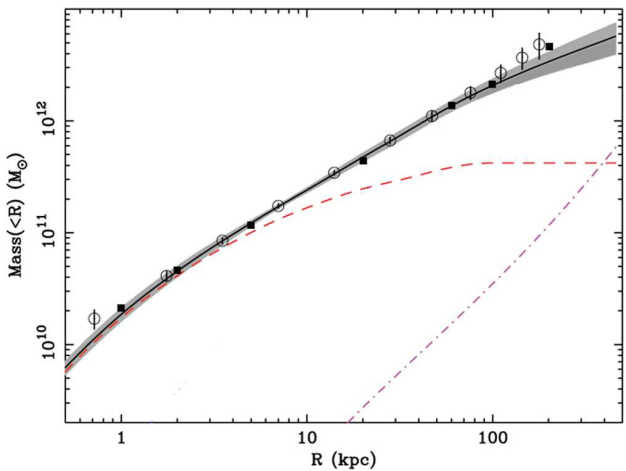}
\caption{NGC 1521: MOND predictions for the enclosed dynamical masses for several radii (filled squares), shown in comparison with the dynamical masses deduced by Ref. \cite{humphrey12}. Except for the MOND points all the rest are reproduced from Ref. \cite{humphrey12}: Gray region, with its central solid (black) line is the range of dynamical masses based on a model of the temperature and x-ray density profiles. The open circles are from their alternative derivation of the masses based on a ``more traditional `smoothed inversion' approach''.
The dashed (red) line is the contribution of the stars $M^*(R)$, for their best fit $M^*/L\_I=2.55\mlsun$, and the dot-dash (blue) line is that of the x-ray gas, $M_g(R)$.  The MOND predictions are given by eq.(\ref{iv}), with $M_b(R)=M^*(R)+M_g(R)$, and the interpolating function $\n_1$.}\label{fig1}
\end{center}
\end{figure}
\begin{figure}
\begin{center}
\includegraphics[width=0.95\columnwidth]{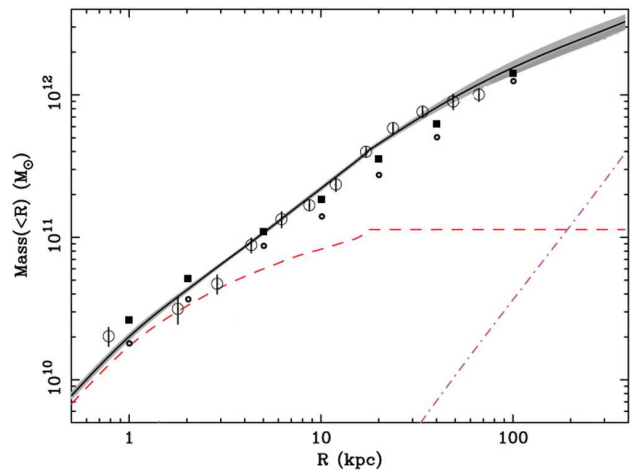}
\caption{NGC 720: The same as Fig.\ref{fig1} (with results from Ref. \cite{humphrey11}), for two sets of MOND predictions: one for the same $M^*/L\_K=0.54\mlsun$ preferred by the Newtonian analysis of Ref. \cite{humphrey11} (small rings); the other for $M^*/L\_K=0.8\mlsun$, which MOND prefers (filled squares). In the latter case the curve for $M^*(r)$ should be raised by a factor 1.5.}\label{fig2}
\end{center}
\end{figure}

In Figures \ref{fig1} and \ref{fig2}, I reproduce from \cite{humphrey11,humphrey12}, the runs of $M^*(R)$, the enclosed stellar mass, $M_g(R)$, the modeled gas mass, and $M_d(R)$ deduced for NGC 1521 and NGC 720, respectively. The stellar contribution is for a spherically averaged triaxial model of the deprojected  light distribution. The conversion to stellar-mass runs assumes their best-fit values of the stellar mass-to-light ratios, which they get from a multi-component fit to $M_d(R)$ (stars + gas + a dark-matter halo): the $I$-band ratio $M^*/L\_I=2.55\mlsun$ for NGC 1521 (with an adopted distance of 62.1\Mpc), and the $K$-band ratio $M^*/L\_K=0.54\mlsun$ for NGC 720 (adopted distance of 25.7\Mpc). For $M_d(R)$,
\cite{humphrey11,humphrey12} give two alternative runs: One derived from a model of  the x-ray temperature and density profiles (extrapolated beyond the measured region). The other run of deduced $M_d$ they describe as based on ``a more traditional `smoothed inversion' approach''.
\par
The MOND predictions of $M_d(R)$, from eq.(\ref{iv}), are shown in the figures, as points, for several values of $R$, calculated using $M_b(R)=M^*(R)+M_g(R)$ .
\par
I take $\az=1.2\cdot10^{-8}\cmss$ (see Ref. \cite{fm12}).
The interpolating function I use is $\m(x)=\m_1(x)=x/(1+x)$, for which $\n_1(y)=(1/2)+(1/4+1/y)^{1/2}$, which was shown in many previous studies \cite{fb05,zf06,sn07,ms08,chiu08,weijmans08} to perform well in the range of x values I cover here. The fact that $\n_1$ is not compatible with solar system constraints for very large $y$ ($y>10^6$) is of no concern here, since I use this form only up to $y\sim 20$ (as in all the above references). For example, using, instead, $\n(y)=\bar\nu\_{1/2}(y)\equiv (1-e^{-y^{1/2}})^{-1}$ (from the families defined in Ref. \cite{ms08}), which is used all along in Ref. \cite{fm12} for MOND rotation curve predictions, would have given the same results here: Despite their different analytic form,  $\bar\nu\_{1/2}$ and $\n_1$ are everywhere equal to better than 4.5\%, while $\bar\nu\_{1/2}$ has the added advantage that it approaches $1$ fast at high $y$; so, unlike $\nu_1$, is consistent with solar system constraints.\footnote{More generally, different interpolating functions coincide at high and low values of their argument; so we expect some differences only in the intermediate-acceleration regime, between about $10 \kpc$ and a few tens $\kpc$.}
\par
The MOND predictions shown in Fig.\ref{fig1} use the same $M^*/L\_I$ value as adopted by Ref. \cite{humphrey12}.
We see that in this case, the MOND predictions agree very well with $M_d(R)$ deduced by Ref. \cite{humphrey12}.
For NGC 720, the best fit $M^*/L\_K=0.54\mlsun$ found by Ref. \cite{humphrey11} gives a MOND-predicted $M_d(R)$ that is somewhat lower than the deduced run. With the system parameters adopted by Ref. \cite{humphrey11}, MOND prefers a higher value of $M^*/L\_K$. Without attempting a best-fit procedure (which would not be more informative, anyway) I also show in Fig.\ref{fig2} the MOND predictions for $M^*/L\_K=0.8\mlsun$. For this value the agreement is good--as good as that between the two alternative runs of $M_d$ presented by Ref. \cite{humphrey11}. This higher value of $M^*/L\_K$ is also in much better agreement with values found in earlier, rotation-curve analyses (see Fig. 2 of Ref. \cite{sv98}, and Fig. 28 of Ref. \cite{fm12}).

\section{\label{discussion}Discussion}
To better appreciate the import of the results, it is helpful to break the full curve comparison into the several salient MOND sub-predictions: (i) In the inner regions, $R\lesssim 10\kpc$, the measured accelerations are rather larger than $\az$ [$g(10\kpc)\approx 3\az,~2.3\az$ in NGC 1521, and NGC 720, respectively, and higher for smaller $R$]; so, MOND predicts only a small mass discrepancy there: it predicts that with a reasonable $M^*/L$ value the runs of $M_d$ and $M_b$ can be made approximately equal in this region; which is indeed the case. (ii) MOND predicts that, with the preferred $M^*/L$ value, the mass discrepancy should set in where $g=GM_d(R)/R^2\sim\az$. Indeed, this is where the discrepancy becomes $\sim 2$. (iii) With the same $M^*/L$ value, MOND predicts correctly the run of the mass discrepancy at large radii, where the accelerations become much smaller than $\az$; the predicted discrepancy is $M_d/M_b\approx (\az/\gN)^{1/2}$. Note that $\az$ appears in (ii) and (iii) in different and independent roles; the fact that the same value matches both is nontrivial.
\par
The two ellipticals considered here differ greatly from the extensively studied disk galaxies, in structure, composition (relative contributions of stars, neutral-gas, and hot-gas), and formation history. Also, their fields were probed by a different method to that of disk galaxies. Hence, the significance of the above results for the MOND-dark-matter controversy goes far beyond the mere evidence that two more galaxies conform with the predictions of MOND. In particular, in the context of the dark-matter paradigm it is all the more unexpected that the relation between baryons and dark matter, found here for the two ellipticals, is described so accurately by the same formula that accounts for disc-galaxy dynamics.
\par
With all the successes of MOND in accounting for the dynamics of galaxies of all types without dark matter, we should remember that MOND does not explain away completely the mass discrepancy in galaxy clusters, where even in MOND some amount of yet undetected matter is required (see Ref. \cite{fm12} for details). Finding this extra mass, or modifying MOND to account for the remaining discrepancy, is a standing challenge for MOND.
\par
The finer details of the comparison of the MOND predictions with the deduced $M_d$ runs may be subject to remaining uncertainties (in the distances, the exact value of $\az$, systematics, etc.).
I want to discuss, specifically, the inner regions of NGC 720, where, with the $M/L$ value preferred by MOND, the MOND prediction is somewhat above the dynamical masses deduced by Ref. \cite{humphrey11}.
\par
In the Newtonian fitting procedure of Ref. \cite{humphrey11}, the best-fit $M^*/L$ value is constrained, as they explain, mainly by their deduced $M_d$ profile in the inner several $\kpc$. At large radii, $M_d$ is dominated by the dark-matter halo, which comes with its own adjustable parameters. But, it is just these inner regions where the x-ray-deduced $M_d$ is most susceptible to various systematics (as discussed in Ref. \cite{humphrey11}). For example, it is there that departures from the assumed hydrostatic equilibrium (such as rotation of the gas, turbulence, infall, etc.) are most likely. In this connection, the extensive study of Ref. \cite{diehl07}, based on a large sample of x-ray ellipticals, is most revealing. They find that in the inner parts of ellipticals, where the field is dominated by the stellar mass component, there is, generally, little correlation between the optical and x-ray morphologies. They conclude that, as a rule, ``the gas is at least so far out of equilibrium that it does not retain any information about the shape of the potential, and that X-ray-derived radial mass profiles may be in error by factors of order unity''.\footnote{Also, departures from spherical symmetry due to the marked ellipticity of stellar mass are most important there.} This view is strengthened by the work of Ref. \cite{ciotti04}, and the recent \cite{humphrey12a}.
\par
In contradistinction, in MOND, $M^*/L$ is the only ``free'' parameter that characterizes the mass distribution. In the whole region analyzed here, the gas mass is subdominant; so, in MOND, $M^*/L$ is constrained by the whole region of study. So, arguably, more weight should be given to the performance of MOND in the regions beyond several $\kpc$, and the tension in the very inner regions of NGC 720 could be easily attributed to departures from equilibrium, as found to be ubiquitous by Ref. \cite{diehl07}. So the MOND $M^*/L$ value appears more reliable than the value determined by Ref. \cite{humphrey11}, which is 33 percent lower. In fact, if we permit ourselves even a larger discrepancy in the inner few $\kpc$, allowing us an even larger $M^*/L$ value, based on the inadequacy of the x-ray mass determination in the inner region, the agreement of the MOND predictions with the measure $M_d$ can be improved in the outer regions.
NGC 1521 seems to avoid this tension in the inner regions, even assuming full equilibrium.
\par
The accuracy of the MOND predictions of the x-ray-deduced fields, over the large range beyond a few $\kpc$ in NGC 720, and over the whole range in NGC 1521, may be taken as evidence that, by and large, the assumption of hydrostatic equilibrium is not strongly amiss there. Otherwise it would require too much of a conspiracy for MOND to correctly predict fictitious $M_d$ runs.
\par
Interestingly, x-ray observations of the inner regions of NGC 720 were claimed, in the past, to evince tension with MOND: It was claimed in Ref. \cite{buote94}, based on ROSAT data, that their x-ray isophote ellipticities were too large to be compatible with the MOND field of the stellar component alone. Later, it became evident that these pronounced ellipticities were largely an artifact of point source contamination. Cleaner Chandra data \cite{buote02} showed rather smaller x-ray ellipticities. But Ref. \cite{buote02} still maintained that even the remaining ellipticiies (within $\sim 18\kpc$), and the degree of misalignment they show with respect to the optical isophotes evince the presence of a dark matter halo, and--without actually giving any calculation in MOND-that a theory like MOND cannot account for these. I think that this should be of little concern for MOND in light of the finding of Ref. \cite{diehl07}. The case for concluding so is even strengthened by what I find in this paper.

I thank Ofer Yaron for help with the figures.

\clearpage
\end{document}